\documentclass[pra,showpacs,showkeys,amsfonts,amsmath]{revtex4}
\usepackage{bm}
\usepackage{graphicx}
\RequirePackage{mathptm}
\newtheorem{thm}{Theorem}[section]

\newtheorem{lem}{Lemma}[section]

\newtheorem{rem}{Remark}[section]
\newtheorem{prop}{Proposition}[section]
\newtheorem{conc}{Conclusion}[section]
\numberwithin{equation}{section}
\newcommand{\si}[1]{\sigma_{#1}}

\newcommand{\ip}[2]{\langle \,{#1},\,{#2}\,\rangle}
\newcommand{\W}[4]{\begin{cases}
#1 ,&#2\\[2mm]
#3 ,&#4
\end{cases}}
\newcommand{\ro}{\varrho}

\newcommand{\la}{\lambda}
\newcommand{\las}[1]{\lambda_{#1}}

\newcommand{\al}{\alpha}
\newcommand{\hla}[1]{\widetilde{\lambda}_{#1}}

\newcommand{\ee}{\varepsilon}
\newcommand{\ep}{\epsilon}

\newcommand{\g}{\gamma}

\newcommand{\om}{\omega}

\newcommand{\te}{\theta}
\newcommand{\vf}{\varphi}

\newcommand{\I}{\openone}
\newcommand{\conj}[1]{\overline{#1}}
\newcommand{\fala}[1]{\widetilde{#1}}
\newcommand{\ket}[1]{|{#1}\rangle}
\newcommand{\bra}[1]{\langle {#1} |}

\newcommand{\C}{\mathbb C}
\newcommand{\R}{\mathbb R}
\newcommand{\M}{\mathbb M}

\newcommand{\D}{\mathbb D}
\newcommand{\PP}{\mathbb P}
\newcommand{\Sp}{\mathbb S}

\newcommand{\fE}{\mathcal E}
\newcommand{\cA}{{\mathcal A}}
\newcommand{\cB}{{\mathcal B}}
\newcommand{\cM}{\mathcal M}
\newcommand{\cP}{\mathcal P}
\newcommand{\cO}{\mathcal O}

\newcommand{\tr}{\mathrm{tr}\,}
\newcommand{\adj}{\mathrm{adj}\,}
\newcommand{\ptr}[1]{\mathrm{tr}_{#1}}

\newcommand{\mr}[1]{\mathrm{#1}}

\newcommand{\DS}{\displaystyle}

\begin{document}
\title{Two-qubit trace-norm geometric discord: the complete solution}
\author{Piotr
{\L}ugiewicz\footnote{piotr.lugiewicz@ift.uni.wroc.pl},  Andrzej Frydryszak \footnote{ andrzej.frydryszak@ift.uni.wroc.pl} and Lech Jak{\'o}bczyk
\footnote{ lech.jakobczyk@ift.uni.wroc.pl} }
 \affiliation{Institute of Theoretical Physics\\ University of
Wroc{\l}aw\\
Plac Maxa Borna 9, 50-204 Wroc{\l}aw, Poland}
\begin{abstract}
We present the complete solution of the problem of determination of trace-norm geometric discord for arbitrary two-qubit state. Final answer is achieved due to effective reduction of the problem to the study of critical points of certain mapping depending on projectors. Our results are illustrated on various, also new, families of two-qubit states and  compared to  already known special solutions.    
\end{abstract}
 \maketitle
\section{Introduction}
It is a common belief that question of the explicit quantitative characterization of correlations for a two-qubit systems can
always be answered, but actually in many cases we land on a uncharted territory, where the explicit answer is not known \cite{Modi, Adesso, HHWPZF}. One of these situations is  the strict solution of the question of trace-norm quantum geometric discord  $D_{1}$ for an arbitrary
state of the two-qubit system. The two-qubit system is the simplest nontrivial compound system which we have at our
disposal to study quantum correlations and all relevant tools invented to measure them \cite{Modi}. The geometric quantum discord
measured by means of the distance induced by the trace-norm seems to be one of the solid candidates. Contrary to the
geometric discord defined by means of the Hilbert-Schmidt distance, it is well defined and satisfies requirements of the
bona fide measure of quantum correlations. As it is well known, the Hilbert-Schmidt distance yields to an anomalous
behavior of related to it geometric discord $D_{2}$, but it has an attractive property, that  it can be
calculated relatively easy. The procedure for calculating geometric discord involves the minimization which for generic
states for the
trace-distance is rather challenging. As we have shown for qutrits and higher dimensional bipartite systems
can hardly be operationally/analytically performed, except for selected families of states \cite{LFJ}.

In the present work we show how explicitly  the $D_{1}$ can be determined, we reveal the geometry of
the problem of minimization, what in turn deepens the understanding of the intrinsic geometry of the two-qubit quantum
state space. For the  Hilbert-Schmidt distance quantum discord $D_{2}$ such a question has been
answered in a compact form by Daki\'{c} et al. in 2010 \cite{GQD}, namely
the explicit formula for the $D_{2}$ has been obtained. The ease of computation does not heal the nonphysical behavior of the $D_{2}$ under the  local evolution of the system \cite{lj-hs} and its non-contractibility under completely positive trace preserving mappings.

The trace-norm measurement-induced geometric discord is well defined correlation measure. For two qubits it is also equal to the standard geometric discord defined as the distance from a given state to the set of classical-quantum states \cite{Na, RSI} and  it is more convenient to use quantity based on the disturbance induced by measurement. Such quantity will be simply called geometric quantum discord. However,
its calculation is still rather challenging, and by now  no explicit  answer for arbitrary two-qubit state has been known.
According to our best knowledge supported by the recent review on the quantum correlations and geometric quantum
discord \cite{HHWPZF}, prior to the present work only few answers were found for selected families of states:
\begin{itemize}
\item[(a)] for the Bell diagonal states (or states with maximally mixed marginals) \cite{Paula},
\item[(b)] for the X-shaped states \cite{cicca},
\item[(c)] for the states with the correlation matrix having only one non-zero singular value and arbitrary Bell vectors of the marginals -- such class contains in particular so called quantum-classical states \cite{cicca}.
\end{itemize}
The objective of this study is to provide the complete solution of the problem of giving the explicit value of
 geometric quantum discord $D_{1}(\rho)$ for arbitrary two-qubit state $\rho$. We find the answer to the minimization
problem which has nice geometric meaning. As an illustration, we analyse previously known partial solutions for mentioned
above  families of states  and indicate mechanism making that these solutions were so much simpler.

The paper is organized as follows. In Section 2, we briefly review basic notions and fix the relevant notation. Then, in
Section 3, we prove the theorem giving characterization of the central formula for the trace-norm of the disturbance of the
system by a local projective measurements. To guide the reader gently to the full solution, in Section 4, we compute
the $D_{1}(\rho)$ for the relatively large family of states for which the minimization procedure is not necessary at all or
is straightforward. The Section 5 is central for the final result, there we study the critical points of the
trace-norm disturbance mapping and obtain that there are two possibilities: singular critical points and  smooth critical
points. As a preliminary application we apply new results to the states with maximally mixed marginals. In the Section 6 we
prove the main result: the determination of the trace norm geometric discord in general case. Finally, the Section 7 will
allow reader to make contact with known solutions and go beyond the family of  the X- shaped states. We finish the
paper with some conclusions and comments on the main findings and obstructions to their extensions to higher
dimensional systems.
\section{Preliminary notions}
In the description of two-qubit systems we follow general formulation of $d$ - level quantum systems
introduced in \cite{LFJ}, specified to the case $d=2$.
\subsection{Two qubits}
Qubit is a $2$- level quantum system. The corresponding Hilbert space equals to $\C^{2}$ and the observables are given by hermitian elements of full
matrix algebra $\M_{2}(\C)$. As a basis in $\M_{2}(\C)$ we use standard Pauli matrices $\si{1},\, \si{2},\, \si{3}$ and the identity matrix $\I_{2}$.
In the following we will use the notation
$$
\ip{n}{\sigma}=n_{1}\si{1}+n_{2}\si{2}+n_{3}\si{3},\quad \sigma=(\si{1},\si{2},\si{3}),\quad n=(n_{1},n_{2},n_{2})
$$
Then from the properties of Pauli matrices it follows that
\begin{equation}
\ip{n}{\sigma}\,\ip{m}{\sigma}=\ip{n}{m}\,\I_{2}+i\,\ip{n\times m}{\sigma}
\end{equation}
The set $\fE_{2}$ of all states of $2$ - level system can be parametrized as follows
\begin{equation}
\ro=\frac{1}{2}\,\left(\I_{2}+\ip{n}{\sigma}\right),\quad n\in \R^{3}\label{stan}
\end{equation}
where $||n||\leq 1$. So $\fE_{2}$ is given by the unit ball in $\R^{3}$ and the pure states correspond to the unit sphere $||n||=1$.
\par
Consider now two qubits $\cA$ and $\cB$. It is convenient to parametrize the set of states of composite system as follows
\begin{equation}
\ro=\frac{1}{4}\,\left(\I_{2}\otimes \I_{2}+\ip{x}{\sigma}\otimes \I_{2}+\I_{2}\otimes \ip{y}{\sigma}+
\sum\limits_{j,k=1}^{3}K_{jk}\,\si{j}\otimes\si{k}\right)\label{stan2}
\end{equation}
where $x,\, y\in \R^{3}$ and $K=(K_{jk})$ is the correlation matrix. Notice that
$$
x_{j}=\tr\,(\ro\,\si{j}\otimes \I_{2}),\quad
y_{j}=\tr\,(\ro\,\I_{2}\otimes \si{j}),\quad K_{jk}=\tr\,(\ro\,
\si{j}\otimes \si{k})
$$
The parametrization (\ref{stan2}) is chosen in such a way that the
marginals $\ptr{\cA}\ro$ and $\ptr{\cB}\ro$ are given by the vectors
$x$ and $y$ as in (\ref{stan}). \par Let us discuss now the adjoint
representation of the group $\mr{SU(2)}$. Let $\hat{U}\in
\mr{SU(2)}$ and define $3\times 3$ matrix $U$ by
$$
\ip{Um}{\sigma}=\hat{U}\,\ip{m}{\sigma}\hat{U}^{\ast}
$$
The matrix $U$ is real and orthogonal. In this way to each
$\hat{U}\in \mr{SU(2)}$ there corresponds $U\in \mr{SO(3)}$. In
contrast to general case, when $d=2$ such obtained group is exactly
equal to the group $\mr{SO(3)}$. Consider now the local
transformations of the state of two qubits
\begin{equation}
\ro \to \hat{U}^{\ast}\otimes \hat{V}^{\ast}\,\ro\, \hat{U}\otimes
\hat{V},\quad \hat{U},\, \hat{V}\in \mr{SU(2)}\label{loc}
\end{equation}
The corresponding vectors $x,\, y$ and the correlation matrix $K$ transform as follows
\begin{equation}
x\to x^{\prime}=Ux,\quad y\to y^{\prime}=Vy,\quad K\to K^{\prime}=UKV^{\mr{T}}
\end{equation}
where $U,\, V$ are the adjoint representations of $\hat{U},\,
\hat{V}$ respectively. Since in the case of qubits we can use the
full group of orthogonal transformations to diagonalize correlation
matrix, any two-qubit state is locally equivalent to the state
with diagonal $K$.
\subsection{Trace-norm geometric discord}
Let $\ro$ be a state of bipartite system $\cA\cB$. When we perform local measurement on the subsystem $\cA$, the state $\ro$ may be disturbed
due to such measurement. The trace-norm (one-sided) measurement induced geometric discord is defined as the minimal disturbance induced by projective measurement $\PP_{\cA}$ on subsystem $\cA$, computed using the trace distance in the set of states. It can be compared with the standard
geometric discord equal to the distance from a given state to the set of classical - quantum states \cite{GQD}. At it was already stated in the Introduction, in the case of qubits these two notions coincide,  and it is more convenient to   use   the quantity based on the disturbance induced by the measurement which will be simply called trace - norm geometric discord. The formal definition is as follows \cite{Paula}
\begin{equation}
D_{1}(\ro)=\min\limits_{\PP_{\cA}}\,||\ro - \PP_{\cA}(\ro)||_{1} \label{discord}
\end{equation}
where $||A||_{1}= \tr\,|A|$.
\par
In the case of qubits, the local projective measurement $\PP_{\cA}$ is given by the one - dimensional projectors $P_{1},\, P_{2}$ on $\C^{2}$, such that
$$
P_{1}+P_{2}=\I_{2},\quad P_{j}P_{k}=\delta_{jk}P_{k}
$$
and $\PP_{\cA}=\PP\otimes \mr{id}$, where
\begin{equation}
\PP(A)=P_{1}AP_{1}+P_{2}AP_{2}\label{proj}
\end{equation}
One - dimensional projectors $P_{k}$ can be always chosen as
$$
P_{k}=uP_{k}^{0}u^{\ast}\quad\text{for some}\quad u\in \mr{SU(2)}
$$
where
$$
P_{1}^{0}=\begin{pmatrix}1&0\\0&0\end{pmatrix},\quad P_{2}^{0}=\begin{pmatrix}0&0\\0&1\end{pmatrix}
$$
Define now a real orthogonal projector $\cP$ on $\R^{3}$
\begin{equation}
\ip{\cP m}{\sigma}=\PP(\ip{m}{\sigma}),\quad m\in \R^{3}
\end{equation}
If $\cP_{0}$ denotes such projector given by (\ref{proj}), where we take $P_{1}^{0}$ and $P_{2}^{0}$, then
$$
\cP_{0}=\mr{diag}(0,0,1)
$$
and
\begin{equation}
\cP=V\cP_{0}V^{\mr{T}},\quad V\in \mr{SO(3)}\label{P}
\end{equation}
Define also orthogonal complements to $\cP_{0}$ and $\cP$
\begin{equation}
\cM_{0}=\I_{3}-\cP_{0},\quad \cM=\I_{3}-\cP
\end{equation}
Obviously $\cM_{0}=\mr{diag}(1,1,0)$,
$$
\cM=V\cM_{0}V^{\mr{T}}, \quad V\in \mr{SO(3)}
$$
and
$$
\mr{dim\, Ran}\, \cM_{0}=\mr{dim\, Ran}\,\cM=2
$$
Notice that in this case the projectors $\cM$ run over the whole set of projectors with dimension $2$.
\par
Now we compute the disturbance of the state (\ref{stan2}) caused by measurement $\PP_{\cA}$. We have
\begin{equation}
S(\cM)=\ro-\PP_{\cA}(\ro)=\frac{1}{4}\left(\ip{\cM x}{\sigma}\otimes \I_{2}+\sum\limits_{k=1}^{3}\ip{\cM K e_{k}}{\sigma}\otimes \ip{e_{k}}{\sigma}\right)
\end{equation}
where $e_{k},\, k=1,2,3$ are the vector of the canonical basis of $\R^{3}$. So
\begin{equation}
D_{1}(\ro)=\min\limits_{\cM}\tr\,|S(\cM)|=\min\limits_{\cM}\tr\,\sqrt{Q(\cM)}
\end{equation}
where $Q(\cM)=S(\cM)S(\cM)^{\ast}$ and the minimum is taken over all projectors $\cM$ on two dimensional subspaces of $\R^{3}$.
\begin{rem}
Notice that the quantity $D_{1}(\ro)$ does not depend on the choice of the state within the class of locally equivalent states. In other words
$$
\min\limits_{\cM}\tr\,\sqrt{Q(\cM)}=\min\limits_{\cM}\tr\,\sqrt{Q^{\prime}(\cM)}
$$
where $Q^{\prime}$ is obtained from $Q$ by taking vector $x^{\prime}=Ux$ and matrix $K^{\prime}=UKV^{\mr{T}}$ instead of $x$ and $K$.
\end{rem}
\section{Trace norm of disturbance $S(\cM)$}
In this section we find the elegant formula for the trace norm of the disturbance $S(\cM)$ (see \cite{cicca} for the other version of such formula).
\begin{thm}
The trace norm of the disturbance $S(\cM)$ of the state (\ref{stan2}) is given by the formula
\begin{equation}
||S(\cM)||_{1}=\frac{1}{\sqrt{2}}\sqrt{||\cM x||^{2}+\tr\,(\cM KK^{\mr{T}})+\sqrt{\left[||\cM x||^{2}+\tr\,(\cM KK^{\mr{T}})\right]^{2}-4\,\left[||K^{\mr{T}}\cM x||^{2}
+\tr\,(E^{\mr{T}}E- \cM E^{\mr{T}}E)\right]}}\label{tracenorm}
\end{equation}
where $||\cdot ||$ denotes the euclidian norm in $\R^{3}$ and $E=\mr{adj}\, K$ is the adjunct matrix of the correlation matrix $K$ (i. e. the transpose of its cofactor matrix).
\end{thm}
\textit{Proof:}
We start the proof of this theorem with the formula for $Q(\cM)=S(\cM)S(\cM)^{\star}$. By a direct computation we obtain
\begin{equation}
Q(\cM)=\frac{1}{16}\left((\tr (\cM KK^{\mr{T}})+\ip{\cM x}{x})\I_{2}\otimes \I_{2}+\I_{2}\otimes 2\ip{K^{\mr{T}}\cM x}{\sigma}+
\sum\limits_{j,k}\tr (K^{\mr{T}}\cM F_{j}\cM K F_{k})\si{j}\otimes \si{k}\right)\label{qM}
\end{equation}
where $F_{k}$ are the generators of $\mr{SO(3)}$, given by
$$
(F_{j})_{kl}=-\ee_{jkl},\quad j,k,l =1,2,3
$$
which satisfy covariance relations
$$
U^{\mr{T}}F_{j}U=\sum\limits_{k}U_{jk}F_{k}
$$
To further simplify (\ref{qM}) consider locally equivalent state for which the correlation matrix is diagonal. So, for a proper transformations $U_{0},\, V_{0}\in \mr{SO(3)}$, we have
\begin{equation}
x_{0}=U_{0}x\quad \text{and}\quad I_{0}=U_{0}KV_{0}^{\mr{T}}\label{loctrans}
\end{equation}
and $I_{0}$ is diagonal
$$
I_{0}=\mr{diag}\, (i_{1},i_{2},i_{3})
$$
Let $E_{0}=\mr{adj} (I_{0})$ be the adjunct matrix of $I_{0}$. Then
$$
E_{0}=\mr{diag} (\ep_{1},\ep_{2},\ep_{3})
$$
where $\ep_{1}=i_{2}i_{3},\, \ep_{2}=i_{1} i_{3},\ep_{3}=i_{1}i_{2}$.  First we show the identity
\begin{equation}
\tr (I_{0}\cM F_{j}\cM I_{0} F_{k})=-2 \ep_{k}V_{j3}V_{k3} \label{identity}
\end{equation}
where $\cM=V\cM_{0}V^{\mr{T}},\, V\in \mr{SO(3)}$. The formula (\ref{identity}) follows since
$$
I_{0}F_{j}I_{0}=\ep_{j}F_{j},\quad j=1,2,3
$$
and
$$
M_{0}F_{k}M_{0}=0,\quad k=1,2 \quad\text{whereas}\quad
M_{0}F_{3}M_{0}=F_{3}
$$
For any orthogonal transformation $V$ define the mapping $\tau_{V}\,:\, \M_{2}(\C)\to \M_{2}(\C)$ as follows
$$
\tau_{V}(a\I_{2}+\ip{m}{\sigma})=a\I_{2}+\ip{Vm}{\sigma}
$$
Now using the identity (\ref{identity}) we can transform the formula (\ref{qM}) to obtain
\begin{equation}
Q(\cM)=\frac{1}{16}\left((\tr (MI_{0}^{2})+\ip{\cM x_{0}}{x_{0}})\I_{2}\otimes \I_{2}+\I_{2}\otimes 2\ip{I_{0}\cM x_{0}}{\sigma}-
2\tau_{V}(\si{3})\otimes \tau_{E_{0}V}(\si{3})\right)\label{qM1}
\end{equation}
Observe that
$$
\tau_{V}(\si{3})=\ip{Ve_{3}}{\sigma}=\ip{v}{\sigma}
$$
where vector $v$ is given by the third column of the matrix $V$. Similarly
$$
\tau_{E_{0}V}(\si{3})=\ip{E_{0}v}{\sigma}
$$
Let us now introduce the matrix
\begin{equation}
R_{0}(\cM)=\I_{2}\otimes \ip{I_{0}\cM x_{0}}{\sigma}-\si{3}\otimes \ip{E_{0}v}{\sigma}
\end{equation}
This matrix has a block diagonal form with  blocks
\begin{equation}
\ip{I_{0}\cM x_{0}-E_{0}v}{\sigma} \quad \text{and}\quad \ip{I_{0}\cM x_{0}+E_{0}v}{\sigma}
\end{equation}
Put
$$
R(\cM)=(\tau_{V}\otimes \mr{id})\,R_{0}(\cM)
$$
then
$$
Q(\cM)=\frac{1}{16}\left( (\tr (\cM I_{0}^{2})+\ip{\cM x_{0}}{x_{0}})\,\I_{2}\otimes \I_{2}+2\, R(\cM)\right)
$$
The spectral analysis of $R_{0}(\cM)$ is easy since characteristic polynomials for the blocks are given by
$$
w_{\pm}(\lambda)=\det\, \left(\ip{I_{0}\cM x_{0}\pm E_{0}v}{\sigma}-\lambda \I_{2}\right) =\lambda^{2}-||I_{0}\cM x_{0}\pm E_{0}v||^{2}
$$
where the $ + (-)$ sign corresponds to the lower (upper) block. One can also check that the vectors $I_{0}\cM x_{0}$ and $E_{0}v$ are orthogonal, so
$$
w_{+}(\lambda)=w_{-}(\lambda)=\lambda^{2}-||I_{0}\cM x_{0}||^{2}-||E_{0}v||^{2}
$$
and the eigenvalues of $R_{0}(\cM)$ are doubly degenerate
$$
\lambda^{(+)}_{\pm}=\lambda^{(-)}_{\pm}=\pm \, \sqrt{||I_{0}\cM x_{0}||^{2}+||E_{0}v||^{2}}
$$
Thus we get
\begin{equation}
\tr\,\sqrt{Q(\cM)}=\frac{1}{2}\,(\omega_{+}+\omega_{-})\label{trpqm}
\end{equation}
where
$$
\omega_{\pm}=\sqrt{||\cM x_{0}||^{2}+\tr (I_{0}^{2}\cM) \pm 2\,\sqrt{||I_{0}\cM x_{0}||^{2}+||E_{0}v||^{2}}}
$$
The formula (\ref{trpqm}) can be further simplified and we arrive at
\begin{equation}
\tr\, \sqrt{Q(\cM)}=\frac{1}{\sqrt{2}}\,\sqrt{||\cM x_{0}||^{2}+\tr (I_{0}^{2}\cM)+\sqrt{\left[ ||\cM x_{0}||^{2}+\tr (I_{0}^{2}\cM)\right]^{2}-
4\left[||I_{0}\cM x_{0}||^{2}+||E_{0}v||^{2}\right]}}
\end{equation}
Now applying the inverse to the transformation (\ref{loctrans}), we finally obtain the formula (\ref{tracenorm}). $\Box$
\section{Solving the minimization problem in the simple case}
To compute the geometric discord of a given state, we have to find minimum of the mapping $\cM\to ||S(\cM)||_{1}$. In general it is a hard problem and its general solution  will be described in the next sections. Now, let us consider a class of two-qubit states for which the solution of this problem is straightforward. This class (denoted by $\fE_{0}$) contains the states (\ref{stan2}) with arbitrary Bloch vectors $x,\,y$ and the correlation matrix of the form
$$
K=t\,V_{0}
$$
where $V_{0}$ is some orthogonal matrix and $t$  is a real parameter. From general properties of the two-qubit states it follows that $t$ belongs to
the interval $|t|\leq 1$, but the actual value of $t$ depends on the choice of the matrix $V_{0}$ and vectors $x,\, y$. To apply the formula (\ref{tracenorm}), notice that
$$
KK^{\mr{T}}=t^{2}\,V_{0}V_{0}^{\mr{T}}=t^{2}\,\I_{3}
$$
and
$$
(\adj K)^{\mr{T}}\,\adj K=\adj (KK^{\mr{T}})=\adj (t^{2}V_{0}V)=t^{4}\,\adj \I_{3}=t^{4}\,\I_{3}
$$
so
$$
\tr (\cM KK^{\mr{T}})=t^{2}\,\tr \cM=2t^{2},\quad \tr (\cM (\adj K)^{\mr{T}}\adj K)=t^{4}\,\tr \cM=2t^{4}
$$
moreover
$$
\tr ((\adj K)^{\mr{T}}\adj K)= t^{4}\tr \I_{3}=3t^{4},\quad
||K^{\mr{T}}\cM x||=|t| \,||\cM x||
$$
Now the formula (\ref{tracenorm}) gives
\begin{equation*}
\begin{split}
||S(\cM)||_{1}^{2}&=\frac{1}{2}\left(||\cM x||^{2}+2t^{2}+\sqrt{[||\cM x||^{2}+2t^{2}]^{2}-4t^{4}-4t^{2}\, ||\cM x||^{2}}\right)\\
&=\frac{1}{2}\left(||\cM x||^{2}+2t^{2}+\sqrt{||\cM x||^{4}}\right)\\
&=t^{2}+||\cM x||^{2}
\end{split}
\end{equation*}
so
$$
||S(\cM)||_{1}^{2}\geq t^{2}
$$
and minimal value is achieved for such $\cM$ which projects on $x^{\perp}$ in $\R^{3}$. Thus we have
\begin{thm}
For every state $\ro \in \fE_{0}$, geometric discord is given by
$$
D_{1}(\ro)=|t|
$$
\end{thm}
\par
Characterization of the class $\fE_{0}$ is not an easy task.  Partial information we can obtain considering locally equivalent states with diagonal correlation matrix. Therefore let us consider the correlation matrix $K$ of the form
$$
K=t\, I
$$
where $I=\mr{diag} (i_{1},\,i_{2},\,i_{3}),\; i_{1},i_{2},i_{3}=\pm 1$. In the following we restrict our analysis to $I_{1}=\I_{3}$ and $I_{2}=\mr{diag} (1,-1,1)$. In the first case we obtain the family of states
\begin{equation}
\ro_{1}=\frac{1}{4}\,\begin{pmatrix}1+a+t&\conj{w}&\conj{z}&0\\
w&1+b-t&2t&\conj{z}\\
z&2t&1-b-t&\conj{w}\\
0&z&w&1-a+t\end{pmatrix}\label{ro1}
\end{equation}
where $a,b\in \R,\, w,z\in \C$. Similarly, in the second case, we have
\begin{equation}
\ro_{2}=\frac{1}{4}\,\begin{pmatrix}1+a+t&\conj{w}&\conj{z}&2t\\
w&1+b-t&0&\conj{z}\\
z&0&1-b-t&\conj{w}\\
2t&z&w&1-a+t\end{pmatrix}\label{ro2}
\end{equation}
Notice that the relation between parameters $a,\, b,\, w,\,z$ and Bloch vectors $x,\,y$ is as follows
$$
x_{1}=\mr{Re}\, z,\, x_{2}=\mr{Im}\,z, x_{3}=\frac{1}{2}(a+b)
$$
and
$$
y_{1}=\mr{Re}\,w,\, y_{2}=\mr{Im}\,w,\, y_{3}=\frac{1}{2}(a-b)
$$
To obtain more specific information, we must restrict the number of parameters. The simplest is the case when Bloch vectors equal to zero vector.
Then (\ref{ro1}) contains the one - parameter family of Werner states
$$
\ro_{\mr{W}}=\frac{1}{4}\,\begin{pmatrix}1+t&0&0&0\\
0&1-t&2t&0\\
0&2t&1-t&0\\
0&0&0&1+t\end{pmatrix},\quad -1\leq t\leq \frac{1}{3}
$$
On the other hand, (\ref{ro2}) contain the family of isotropic states
$$
\ro_{\mr{iso}}=\frac{1}{4}\,\begin{pmatrix}1+t&0&0&2t\\
0&1-t&0&0\\
0&0&1-t&0\\
2t&0&0&1+t\end{pmatrix},\quad -\frac{1}{3}\leq t \leq 1
$$
Isotropic states can be obtained as a mixture of maximally mixed state $\ro_{\infty}=\frac{1}{4}\I_{4}$ and maximally entangled state $\Psi_{+}$
given by
$$
\Psi_{+}=\frac{1}{\sqrt{2}}\begin{pmatrix}1\\0\\0\\1\end{pmatrix}
$$
i.e.
$$
\ro_{\mr{iso}}=(1-t)\ro_{\infty}+t\, \ket{\Psi_{+}}\bra{\Psi_{+}}
$$
On the other hand, the Werner states $\tilde{\ro}_{\mr{W}}$ with reversed parametrization ($t\to -t$) can be obtained as he mixture of
$\ro_{\infty}$ and the state $\Phi_{-}$, where
$$
\Phi_{-}=\frac{1}{\sqrt{2}}\begin{pmatrix}\hspace*{2mm}0\\\hspace*{2mm}1\\-1\\\hspace*{2mm}0\end{pmatrix}
$$
So both states are locally equivalent.
\par
More interesting is the case of states with  non-trivial Bloch vectors. For simplicity we consider the vectors of the form
$$
x=y=(0,0,\al),\quad |\al|\leq 1
$$
In this case the properties of the states (\ref{ro1}) and (\ref{ro2}) differ significantly. In particular, $\ro_{1}$ is positive - definite when
$$
|\al|\leq \frac{2}{3}\quad \text{and}\quad 2|\al| -1 \leq t\leq \frac{1}{3}
$$
On the other hand, the positivity region for $\ro_{2}$ is given by the conditions
$$
-\frac{1}{3}\leq t\leq 1\quad\text{and}\quad |\al|\leq \frac{1}{2}\,\sqrt{1+2t-3t^{2}}
$$
\section{Critical points of the mapping $\cM\to ||S(\cM)||_{1}$}
\subsection{Formulation of the problem}
In this section we start to analyze the real problem of finding minimum of the trace norm $||S(\cM)||_{1}$ in general case. As a first step
we will study critical points of the mapping $\cM\to ||S(\cM)||_{1}$.  Since the projector $\cP=\I_{3}-\cM$  is one dimensional, there is a unit vector $v\in \R^{3}$ such that $\cP=P_{v}$. The vector $v$ is given by the third column of the matrix $V\in \mr{SO(3)}$ relating $\cP$ and $\cP_{0}$ (\ref{P}). To study critical points of trace norm of disturbance it is useful to consider the auxiliary function $\mathbf{g}$ defined on the unit sphere $\Sp^{2}\subset \R^{3}$ with values in $\R^{2}$.  The function is defined as follows
\begin{equation}
\mathbf{g}(v)=(g_{1}(v),\, g_{2}(v))\label{g}
\end{equation}
where
\begin{equation}
g_{1}(v)=||(\I_{3}-P_{v})x||^{2}+\tr\,((\I_{3}-P_{v})KK^{\mr{T}})\label{g1}
\end{equation}
and
\begin{equation}
g_{2}(v)=4\,\left(||K^{\mr{T}}(\I_{3}-P_{v})x||^{2}+\tr\, (E^{\mr{T}}E\,P_{v})\right)\label{g2}
\end{equation}
Notice that using the functions (\ref{g1}) and (\ref{g2}), the formula (\ref{tracenorm}) can be rewritten as
$$
||S(\cM)||_{1}=||S(\I_{3}-P_{v})||_{1}=\frac{1}{\sqrt{2}}\,\sqrt{g_{1}(v)+\sqrt{g_{1}(v)^{2}-g_{2}(v)}}
$$
In the following we will consider the mapping $\mathbf{g}$ as a function of two-dimensional projectors $\cM$ or vectors $v\in \Sp^{2}$.
\par
The formulas (\ref{g1}) and (\ref{g2}) can be further simplified if we introduce the following operators
\begin{equation}
W_{y}z=\ip{z}{y}y,\quad y,\, z\in \R^{3}\label{Ptylda}
\end{equation}
and
\begin{equation}
L_{+}=KK^{\mr{T}}+W_{x}\label{Lplus}
\end{equation}
where $x$ is a Bloch vector. Now we obtain
\begin{equation}
g_{1}(v)=||x||^{2}-\ip{x}{v}^{2}+\tr\, (KK^{\mr{T}})-\ip{v}{KK^{\mr{T}}v}=
\tr\,L_{+}-\ip{v}{L_{+}v}
\label{g1p}
\end{equation}
and
\begin{equation}
g_{2}(v)=4\,\left(||K^{\mr{T}}(x-\ip{x}{v}x||^{2}+\ip{v}{E^{\mr{T}}Ev}\right)\label{g2p}
\end{equation}
From the equation (\ref{g1p}) we obtain in particular that
\begin{equation}
\min \limits_{v\in \Sp^{2}}g_{1}(v)=\tr\, L_{+}-\lambda_{3}=\lambda_{1}+\lambda_{2}\label{ming1}
\end{equation}
where $\las{1}\leq \las{2}\leq \las{3}$ are the eigenvalues of real non-negative matrix
$L_{+}$. The relation (\ref{ming1}) allows to obtain the first general result
concerning the value of geometric discord $D_{1}(\ro)$. Namely we have
\begin{thm}
For any two-qubit state (\ref{stan2}) we have
\begin{equation}
D_{1}(\ro)\geq\frac{1}{\sqrt{2}}\,\sqrt{\las{1}+\las{2}}\label{lb}
\end{equation}
where $\las{1}\leq \las{2}\leq \las{3}$ are the eigenvalues of
$L_{+}=KK^{\mr{T}}+W_{x}$ .
\end{thm}
From the relation (\ref{ming1}) we can also  derive well known  exact formula for Hilbert - Schmidt norm geometric discord $D_{2}(\ro)$ \cite{GQD}. By definition
$$
D_{2}(\ro)=\min\limits_{\cM} 2\,\tr Q(\cM)
$$
Using (\ref{qM}) we obtain
$$
\tr\, Q(\cM)=\frac{1}{4}\left(\tr\, (\cM KK^{\mr{T}})+\ip{\cM x}{x}\right)=\frac{1}{4}\,g_{1}(v)
$$
so
$$
D_{2}(\ro)=\min\limits_{v\in \Sp^{2}}\frac{1}{2}\,g_{1}(v)=\frac{1}{2}\,(\las{1}+\las{2})
$$
Obviously, for any two-qubit state
$$
D_{1}(\ro)\geq \sqrt{D_{2}(\ro)}
$$
The right hand side of (\ref{lb}) in some cases gives not only the lower bound but the exact value of discord $D_{1}$. It can happen for such states
$\ro$ for which $D_{1}(\ro)=\sqrt{D_{2}(\ro)}$. Below we show it for an explicit family of states. Consider the states \cite{MM}
\begin{equation}
\ro_{\te}=\frac{1}{4}\,\begin{pmatrix}2\,\cos^{2}\te&0&0&\sin 2\te\\
0&0&0&0\\
0&0&2&0\\
\sin 2\te&0&0&2\sin^{2}\te
\end{pmatrix},\quad \te\in [0,\pi/2]\label{rote}
\end{equation}
For the family (\ref{rote}), we have
$$
x=\begin{pmatrix}0\\0\\-\sin^{2}\te\end{pmatrix},\quad y=\begin{pmatrix}0\\0\\\cos^{2}\te\end{pmatrix}
$$
and
$$
K=\mr{diag} (\cos\te \sin\te,\, -\cos\te \sin \te,\, 0)
$$
so
\begin{equation}
KK^{\mr{T}}+W_{x}=\mr{diag} (\cos^{2}\te\sin^{2}\te,\, \cos^{2}\te\sin^{2}\te,\, \sin^{4}\te)\label{Krote}
\end{equation}
The order of eigenvalues of (\ref{Krote}) depends on $\te$. For $\te\in [0,\pi/4]$
$$
\sin^{4}\te \leq\sin^{2}\te\cos^{2}\te
$$
and for such $\te$
$$
\frac{1}{\sqrt{2}}\sqrt{\las{1}+\las{2}}=\frac{1}{\sqrt{2}}\sqrt{\sin^{4}\te+\sin^{2}\te\cos^{2}\te}=
\frac{1}{\sqrt{2}}\sin\te
$$
On the other hand, one can check that
$$
D_{1}(\ro_{\te})=\frac{1}{2}\,\sin 2\te\geq \frac{1}{\sqrt{2}}\,\sin\te
$$
so in this case we obtain only the lower bound. Now for $\te\in (\pi/4,\,\pi/2]$
$$
\sin^{2}\te\cos^{2}\te\leq \sin^{4}\te
$$
and for such $\te$
$$
\frac{1}{\sqrt{2}}\sqrt{\las{1}+\las{2}}=\frac{1}{\sqrt{2}}\sqrt{2\,\sin^{2}\te\cos^{2}\te}=\sin\te\cos\te=\frac{1}{2}\,\sin 2\te
$$
which equals to the value of $D_{1}(\ro_{\te})$.
\subsection{Singular critical points}
We see that the problem of finding minimum of the mapping $\cM\to ||S(\cM)||_{1}$ is relatively easy when it is achieved on the set
\begin{equation}
\D=\{v\in \Sp^{2}\; :\; g_{1}^{2}(v)=g_{2}(v)\} \label{D}
\end{equation}
Each critical point $v_{0}\in \D$ is called \textit{a singular critical point} in contrast to the \textit{smooth critical point} i.e. critical point  belonging to the set $\Sp^{2}\setminus \D$. When the minimum is achieved at some singular critical point, the minimal value of trace norm $||S(\cM)||_{1}$  is given by the minimum of the function $g_{1}$. To study the properties of the set $\D$, consider another representations of the functions $g_{1}$ and $g_{2}$. Let $L_{-}=KK^{\mr{T}}-W_{x}$. One can check that
\begin{equation}
g_{1}(v)=\tr\, L_{-}-\ip{v}{L_{-}v}+2\,(||x||^{2}-\ip{x}{v}^{2})\label{g17}
\end{equation}
and
\begin{equation}
 g_{2}(v)=4\,(||x||^{2}-\ip{x}{v}^{2})\,\left[\tr\,
L_{-}-\ip{v}{L_{-}v}+||x||^{2}-\ip{x}{v}^{2}\right]
+4\ip{v}{\mr{adj}\, L_{-}\,v}\label{g27}
\end{equation}
where we dropped the term proportional to $1-||v||^{2}$. Using
(\ref{g17}) and (\ref{g27}) one can show that the set (\ref{D}) can
be discrete or it can be equal to the whole sphere $\Sp^{2}$. It
follows from the following Lemma, which can be proved by a direct
computation.
\begin{lem}
Let $\hla{1},\, \hla{2},\, \hla{3}$ be the eigenvalues of the matrix $L_{-}$. The vectors $v\in \D$ satisfy
\begin{equation}
(\hla{1}-\hla{2})^{2}\,(v_{3}^{2}-v_{1}^{2}v_{2}^{2})+(\hla{2}-\hla{3})^{2}(v_{1}^{2}-v_{2}^{2}v_{3}^{2})+
(\hla{1}-\hla{3})^{2}(v_{2}^{2}-v_{1}^{2}v_{3}^{2})=0\label{Drownanie}
\end{equation}
\end{lem}
\begin{rem}
One can check that  in the discrete case, the set $\D$ can contain only one or two projectors, if we identify the vectors $v$ and $-v$.
\end{rem}
For the further analysis it is crucial to have a convenient criterion that enables to decide whether the value of $D_{1}(\ro)$ is achieved at a critical point belonging to the set $\D$ or $\Sp^{2}\setminus \D$.
To obtain such criterion, consider the function
$$
F(t,s)=t+\sqrt{t^{2}-s}
$$
defined on the region
$$
\Omega=\{(t,s)\in \R^{2}\; ;\; t\geq 0,\, s\geq 0,\, t^{2}\geq s\}
$$
For any $t_{\ast}>0$ consider the function
$$
\delta_{t_{\ast}}(t)=\W{t^{2}}{t\geq
t_{\ast}}{2t_{\ast}t-t_{\ast}^{2}}{t_{\ast}/2 \leq t\leq t_{\ast}}
$$
And finally, define the family of regions $\Omega_{t_{\ast}},\, t_{\ast}\geq 0$ such that $\Omega_{0}=\Omega$ and for $t_{\ast}>0$
$$
\Omega_{t_{\ast}}=\{ (t,s)\in \R^{2}\; ;\; t\geq 0,\, s\geq 0,\,
\delta_{t_{\ast}}(t)\geq s\}
$$
Suppose now that the function $g_{1}$ restricted to $\D$ achieves at some vector $v_{\ast}\in \D$ its absolute minimum. Let $\mr{Ran}\, \mathbf{g}$ be
the range of the function $\mathbf{g}$. The values of $D_{1}(\ro)$ is achieved at $v_{\ast} \in \D$ if and only if
$$
\mr{Ran}\, \mathbf{g}\subset \Omega_{g_{1}(v_{\ast})}
$$
So we obtain
\begin{prop}
Let the function $g_{1}$ restricted to $\D$ achieves its absolute minimum at the vector $v_{\ast}\in \D$. The value of $D_{1}(\ro)$ is achieved at this vector, if and only if the following conditions are satisfied
\begin{equation}
g_{1}(v_{\ast})\leq 2\, g_{1}(v)\label{war1}
\end{equation}
for all $v\in \Sp^{2}$, and
\begin{equation}
g_{1}(v_{\ast})g_{1}(v)\geq \frac{1}{2}\left(g_{2}(v)+g_{2}(v_{\ast})\right)\label{war2}
\end{equation}
for all $v\in \Sp^{2}$ such that $g_{1}(v)\leq g_{1}(v_{\ast})$.
\end{prop}
\begin{rem}
Notice that if the conditions (\ref{war1}) and (\ref{war2}) are not satisfied, then the value $g_{1}(v_{\ast})$ gives the upper bound for
$D_{1}(\ro)$ i.e.
$$
D_{1}(\ro)\leq \frac{1}{\sqrt{2}}\,\sqrt{g_{1}(v_{\ast})}
$$
\end{rem}
Now we determine the minimal value $g_{1}(v_{\ast})$. To this end, consider the following cases:\\[2mm]
\textbf{1.} $\hla{1}>\hla{2}>\hla{3}$. The solutions of (\ref{Drownanie}) are given by
\begin{equation}
v_{1}^{2}=\frac{\hla{1}-\hla{2}}{\hla{1}-\hla{3}},\quad v_{1}^{2}=0,\quad v_{3}^{2}=\frac{\hla{2}-\hla{3}}{\hla{1}-\hla{3}}\label{Dsolution}
\end{equation}
and the expression for $g_{1}(v),\; v\in \D$ reads
\begin{equation}
g_{1}(v)=2\,\left(\hla{2}+||x||^{2}-\ip{x}{v}^{2}\right)\label{g10}
\end{equation}
Note that $\hla{2}\geq 0$. The equality (\ref{g10}) is valid in any
orthonormal basis. But if we choose the basis of ordered and
normalized eigenvectors of $L_{-}$, then
$$
g_{1}(v_{\ast})=2\,\left(\hla{2}+\frac{1}{\hla{1}-\hla{3}}\left[ (\hla{1}-\hla{3})x_{2}^{2}+\left(\sqrt{\hla{2}-\hla{3}}\, |x_{1}| -\sqrt{\hla{1}-\hla{2}}\,|x_{3}|\right)^{2}\right]\right)
$$
\textbf{2.} $\hla{1}=\hla{2}>\hla{3}$. The equation (\ref{Drownanie}) gives $v_{1}=v_{2}=0$ and $v_{3}^{2}=1$. So (\ref{g10}) is also valid and we obtain
$$
g_{1}(v_{\ast})=2\,(\hla{2}+x_{1}^{2}+x_{2}^{2})
$$
\textbf{3.} $\hla{1}=\hla{2}=\hla{3}$. In this case $\D=\Sp^{2}$ and the formula (\ref{g10}) is valid.\\[2mm]
So we have the following
\begin{prop}
For all vectors $v\in \D$
\begin{equation}
g_{1}(v)=2\,\left(\mr{int}\{ \hla{1},\hla{2},\hla{3}\}+||x||^{2}-\ip{x}{v}^{2}\right)\label{g11}
\end{equation}
where $\mr{int}$ denotes the intermediate value and $\hla{1},\hla{2},\hla{3}$ are the eigenvalues of the matrix $L_{-}$.
\end{prop}
\begin{rem}
Notice that if the Bloch vector $x$ is zero vector, then the function $g_{1}$ is constant on the set $\D$ and equals to $2\, \mr{int}\, \{\las{1},\las{2},\las{3}\}$, where $\las{1},\las{2},\las{2}$ are the eigenvalues of  the matrix $KK^{\mr{T}}$.
\end{rem}
\subsection{Smooth critical points}
Smooth critical points $v_{0}\in \Sp^{2}\setminus \D$ satisfy
$$
g_{1}^{2}(v_{0})\neq g_{2}(v_{0})
$$
Applying differential analysis to the set $\Sp^{2}\setminus \D$ one can find all smooth critical points of the mapping $\cM\to ||S(\cM)||_{1}$. Let
$$
f^{2}=g_{1}+\sqrt{g_{1}^{2}-g_{2}}-\la\,(1-||v||^{2})
$$
with the constrain $||v||=1$ and where $\la$ is a Lagrange multiplier. For the smooth critical points one obtains
\begin{equation}
4\mu\; \mr{grad}\, g_{1}-\mr{grad}\, g_{2}=-8\om\, v\label{sc16}
\end{equation}
where $2\om= \la\,\sqrt{g_{1}^{2}-g_{2}}$. Moreover $4\,\mu=2\,f^{2}$ i.e.
\begin{equation}
||S(\cM)||_{1}=\sqrt{\mu}\label{sc17}
\end{equation}
\begin{rem}
The condition (\ref{sc16}) in particular means that smooth critical points of the mapping $\cM\to ||S(\cM)||_{1}$ are critical points of the function $\mathbf{g}$ defined by (\ref{g}).
\end{rem}
To find critical points of $\mathbf{g}$  we look for the possible solutions of the equation
\begin{equation}
\left[W_{x}P_{v}KK^{\mr{T}}+KK^{\mr{T}}P_{v}W_{x}-KK^{\mr{T}}W_{x}-W_{x}KK^{\mr{T}}+E^{\mr{T}}E+\mu\,L_{+}\,
\right]v=\om\, v\label{sc18}
\end{equation}
A vector $v\in \Sp^{2}$ is a solution of (\ref{sc18})  if there exists a real number $\mu$ such that $v$ is a solution of the above eigenvector
problem. Applying the above results one can prove the following Theorem:
\begin{thm}
Any smooth critical point of the mapping $\cM\to ||S(\cM)||_{1}$ satisfies (\ref{sc18}) i.e. is a critical point of the function $\mathbf{g}$.
\end{thm}
\begin{rem}
One can also show that critical points contained in the set $\D$ satisfy (\ref{sc18}), so the thesis of Theorem V.2 is true for all critical points
of the mapping $\cM\to ||S(\cM)||_{1}$.
\end{rem}
The condition in equation (\ref{sc18}) can be rewritten in the following way. Notice that the two-dimensional projector $\cM$ is a critical point of the function $\mathbf{g}$ if and only if it commutes with the matrix
\begin{equation}
G_{\mu}=-W_{x}\cM KK^{\mr{T}}- KK^{\mr{T}}\cM W_{x}+E^{\mr{T}}E+\mu
L_{+}\label{sc19}
\end{equation}
i.e. $v$ is the solution of (\ref{sc18}) if and only if
\begin{equation}
G_{\mu}P_{v}=P_{v}G_{\mu}\label{sc20}
\end{equation}
On the other hand, the information about the Lagrange multiplier
$\om$ can be recovered from the equation (\ref{sc20}) that leads to
\begin{equation}
G_{\mu}P_{v}=\om\, P_{v}\label{sc21}
\end{equation}
\subsection{First application: states with maximally mixed marginals}
General analysis of the equations (\ref{sc18}) or (\ref{sc20}) will
be presented in the next section, here we consider the first
application to the states with maximally mixed marginals (MMM
states) i.e. such states $\ro$ that
$$
\ptr{\cA}\ro=\ptr{\cB}\ro=\frac{1}{2}\I_{2}
$$
In this case the Bloch vectors $x$ and $y$ are zero vectors. Let $K$ be the correlation matrix of MMM state $\ro$. Using the local transformation one can bring the matrix $KK^{\mr{T}}$ to the diagonal form $\mr{diag}\, (i_{1}^{2},\, i_{2}^{2},\, i_{3}^{2})$ with $i_{1}^{2}\geq i_{2}^{2}\geq i_{3}^{2}$. Now the equation (\ref{Drownanie}) for the set $\D$ has a form
\begin{equation}
(i_{1}^{2}-i_{2}^{2})^{2}\,(v_{3}^{2}-v_{1}^{2}v_{2}^{2})+(i_{2}^{2}-i_{3}^{2})^{2}\,(v_{1}^{2}-v_{2}^{2}v_{3}^{2})+
(i_{1}^{2}-i_{3}^{2})^{2}\,(v_{2}^{2}-v_{1}^{2}v_{3}^{2})=0\label{DMMM}
\end{equation}
Moreover
\begin{equation}
G_{\mu}=E^{\mr{T}}E+\mu\,KK^{\mr{T}}=\mr{diag}\,\left(i_{2}^{2}i_{3}^{2}+\mu\,i_{1}^{2},\, i_{1}^{1}i_{3}^{2}+\mu\, i_{2}^{2},\, i_{1}^{2}i_{2}^{2}+\mu\,
i_{3}^{2}\right)\label{Gmu}
\end{equation}
so the general form of $P_{v}$ satisfying (\ref{sc21}) depends on the degeneracy of the eigenvalues of $G_{\mu}$.\\
Consider first the case when
$i_{1}^{2}>i_{2}^{2}>i_{3}^{2}$. Suppose that $\mu\notin \{i_{1}^{2},\, i_{2}^{2},\, i_{3}^{2}\}$, then the eigenvalues of $G_{\mu}$ are non-degenerate and the only possible solutions of (\ref{sc21}) correspond to the vectors
$$
(1,0,0),\quad (0,1,0)\quad\text{and}\quad (0,0,1)
$$
which, according to (\ref{DMMM}) cannot belong to the set $\D$. So
they can be smooth critical points of $\cM\to ||S(\cM)||_{1}$. The
condition (\ref{sc17}) enforces $\mu=i_{1}^{2}$ or $\mu =
i_{2}^{2}$, which is in a contradiction with the assumption that
$\mu\notin \{i_{1}^{2},\, i_{2}^{2},\, i_{3}^{2}\}$. So we conclude
that the square of the discord can be achieved in the set $\mu\in
\{i_{1}^{2},\, i_{2}^{2}\}$, since $i_{3}^{2}$ can be excluded
because of the lower bound given by Theorem V.1
$$
2\,\mu \geq i_{2}^{2}+i_{3}^{2}
$$
Since $i_{2}^{2}<i_{1}^{2}$, we get
\begin{equation}
D_{1}(\ro)=\mr{int}\, \{|i_{1}|,\, |i_{2}|,\, |i_{3}|\}\label{dMMM}
\end{equation}
Similarly one also obtains the formula (\ref{dMMM}) in the case when $i_{1}^{2}=i_{2}^{2}>i_{3}^{2}$ or $i_{1}^{2}>i_{2}^{2}=i_{3}^{2}$.
The case $i_{1}^{2}=i_{2}^{2}=i_{3}^{2}$ is trivial: the set $\D$ is equal to the whole sphere $\Sp^{2}$ and the function (\ref{sc17}) is constant
and we again obtain (\ref{dMMM}). Thus we confirm the known result (see e.g. \cite{Paula})
\begin{conc}
For any MMM state trace - norm geometric discord is equal to the intermediate value in the set of singular values of the corresponding correlation matrix.
\end{conc}
\section{Main result: determination of quantum discord in general case}
In this section we use the equation
(\ref{sc20}) to solve the main problem, namely to find the value of geometric discord in general case. It is convenient to consider the
matrix (\ref{sc19}) with respect to ordered basis of orthonormal
eigenvectors of the operator $L_{-}$ (we will assume that
$\hla{1}\geq \hla{2}\geq \hla{3}$). When we also neglect the parts
which commute with the projectors $P_{v}$, we obtain the matrix
$$
\fala{G}_{\mu}=W_{x}P_{v}\Lambda
+\Lambda\,P_{v}W_{x}+2\,P_{x}W_{v}P_{x}-(\hla{1}-\hla{2})W_{x}+\fala{\mu}(\Lambda+2\,W_{x})+
(\hla{1}-\hla{2})(\hla{2}-\hla{3})P_{y_{0}}
$$
where
$$
\Lambda=\begin{pmatrix}\hla{1}-\hla{3}&0&0\\0&\hla{2}-\hla{3}&0\\0&0&0\end{pmatrix},\quad y_{0}=\begin{pmatrix}0\\1\\0\end{pmatrix}
$$
and
$$
\fala{\mu}=\mu-\hla{2}-||x||^{2}
$$
In the following we will also use the matrices $\fala{\Lambda}=\Lambda -(\hla{1}-\hla{2})\,\I_{3}$ and
$$
\Pi=\begin{pmatrix}0&(\hla{1}-\hla{2})x_{3}&-(\hla{1}-\hla{3})x_{2}\\[2mm](\hla{1}-\hla{2})x_{3}&0&(\hla{2}-\hla{3})x_{1}\\[2mm]
-(\hla{1}-\hla{3})x_{2}&(\hla{2}-\hla{3})x_{1}&0\end{pmatrix}
$$
Notice that if $L_{-}$ has a degenerate spectrum, the form of the
operator $\fala{G}_{\mu}$ is particulary simple and the desired
minimal value of $\cM\to ||S(\cM)||_{1}$ can be found in a
straightforward way.
\subsection{Non-degenerate spectrum of $L_{-}$}
Assume that $\hla{1}>\hla{2}>\hla{3}$. First we find the location of critical points of the mapping $\cM\to ||S(\cM)||_{1}$. Notice that in the following all components of the vector $v$ and the Bloch vector $x$ are computed with respect to the ordered basis of orthonormal eigenvectors of $L_{-}$.
\begin{prop}
Assume that the matrix $L_{-}=KK^{\mr{T}}-W_{x} $ has non -
degenerate eigenvalues. The critical points of the mapping $\cM\to
||S(\cM)||_{1}$ ($\cM=\I_{3}-P_{v}$) are contained in the subset of
vectors $v\in\Sp^{2}$ satisfying the condition
$$
v_{1}v_{2}v_{3}=0
$$
i.e. in the big circles in $\Sp^{2}$, where the components of $v$
are given with respect to the ordered basis of eigenvectors of
$L_{-}$. In particular the value of $D_{1}(\ro)$ is achieved only on
such vectors.
\end{prop}
\textit{Proof}: Suppose that critical points do not satisfy the
condition $v_{1}v_{2}v_{3}=0$ and moreover let
$v_{3}x_{1}-v_{1}x_{3}\neq 0$. The condition $v_{1}v_{2}v_{3}\neq 0$
implies that the critical points are outside the set $\D$. Notice
that $\fala{\mu}+\ip{x}{v}^{2}\geq 0$. Since (\ref{sc20}) is satisfied, the following strict inequality is true
$$
\fala{\mu}+\ip{x}{v}^{2}>0
$$
By a direct computation one can find the representation of $\fala{\mu}$
$$
(\fala{\mu}+\ip{x}{v}^{2})(v_{3}x_{1}-v_{1}x_{3})=-(v_{3}x_{1}-v_{1}x_{3})\ip{v}{\fala{\Lambda}v}-\frac{1}{2}v_{2}\ip{v}{\Pi v}
$$
along the critical points of the function $\mathbf{g}$, described by
the condition
\begin{equation}
v_{2}\ip{v}{\Pi v}+(v_{3}x_{1}-v_{1}x_{3})\ip{v}{\fala{\Lambda} v}=0\label{cg31a}
\end{equation}
or
\begin{equation}
\ip{x}{v}(v_{3}x_{1}-v_{1}x_{3})+(\hla{1}-\hla{3})v_{1}v_{3}=0\label{cg31b}
\end{equation}
To find the critical points of the mapping $\cM\to ||S(\cM)||_{1}$ we use the condition (\ref{sc17}), which now reads
$$
4(\hla{1}-\hla{2})(\hla{2}-\hla{3})(v_{3}x_{1}-v_{1}x_{3})^{2}v_{2}=\ip{v}{\Pi v}(v_{2}\ip{v}{\Pi v}+2(v_{3}x_{1}-v_{1}x_{3})\ip{v}{\fala{\Lambda} v})
$$
To obtain the result, let us assume that the equation (\ref{cg31a}) has a non-empty set of solutions. Then we get that
$$
2\fala{\mu}=-2\ip{x}{v}^{2}-\ip{v}{\fala{\Lambda} v}
$$
But this representation can not be valid outside the set $\D$ and we obtain the contradiction with our assumption. Let now the equation (\ref{cg31b})
has the non-empty set of solutions. Then we get
$$
(\hla{1}-\hla{3})(\fala{\mu}+\ip{x}{v}^{2})v_{1}v_{3}=(\hla{1}-\hla{3})v_{1}v_{3}(2(\fala{\mu}+\ip{x}{v}^{2})+\ip{v}{\fala{\Lambda} v})
$$
and after simplifications we obtain
$$
\fala{\mu}=-\ip{x}{v}^{2}-\ip{v}{\fala{\Lambda}v}
$$
But this is impossible, as this leads to the condition $v_{2}=0$.
Finally, let $v_{3}x_{1}-v_{1}x_{3}=0$. Then the equation
(\ref{sc20}) gives
$$
\fala{\mu}+\ip{x}{v}^{2}=0
$$
This condition together with (\ref{sc17}) leads again to $v_{2}=0$. $\Box$\\
The assumption of non-degeneracy of the spectrum of $L_{-}$ is relevant in the above result. On the other hand, the proof suggests that the
location of critical points of the mapping $\cM\to ||S(\cM)||_{1}$ given by Proposition VI.1 is not optimal. Indeed, if for example $x_{1}\neq 0$,
the critical points cannot satisfy $v_{1}=0$ if $v_{2}v_{3}\neq 0$. Similarly, if $x_{3}\neq 0$, the critical points never lay on the big circle
$v_{3}=0$ with $v_{1}v_{2}\neq 0$, unless $\hla{1}+\hla{3}>2\, \hla{2}$ and the following condition is satisfied
$$
x_{1}^{2}+x_{2}^{2}=\frac{\hla{1}-\hla{2}}{\hla{2}-\hla{3}}\, x_{1}^{2}
$$
The additional informations of this kind can simplify the determination of quantum discord in some situations. Nevertheless, we will not discussing
this improvements here and focus on the direct method of finding the absolute minimum of the mapping $\cM\to ||S(\cM)||_{1}$.
\par
Now we can go to the solution of the main problem. This is a problem of minimizing the quadratic form $\mu$, restricted to the big circles. The calculations are elementary but tedious, so we do not present all details. \\
1. Let us start with $v_{1}=0$. One can find
$$
\fala{\mu}=\hla{1}-\hla{2}-\ip{x}{v}^{2}
$$
and
$$
\fala{\mu}_{\mr{min}}=\hla{1}-\hla{2}-||x||^{2}+x_{1}^{2}
$$
So
\begin{equation*}
D_{1}(\ro)^{2}\leq \hla{1}+x_{1}^{2}
\end{equation*}
2. Let $v_{3}=0$. Then
$$
\fala{\mu}=(\hla{1}-\hla{2})v_{2}^{2}-\ip{x}{v}^{2}
$$
Inserting the minimum $\fala{\mu}_{\mr{min}}$ one gets
$$
D_{1}(\ro)\leq\frac{1}{\sqrt{2}}\sqrt{\hla{1}+\hla{2}+||x||^{2}+x_{3}^{2}-
\sqrt{(\hla{1}-\hla{2}+||x||^{2}-x_{3}^{2})^{2}-4(\hla{1}-\hla{2})x_{2}^{2}}}
$$
3. Let finally $v_{2}=0$. This case is not so straightforward. One
needs to introduce the additional parameter. We start with the
formula
\begin{equation}
2\fala{\mu}=\hla{1}-\hla{2}-(\hla{1}-\hla{3})v_{1}^{2}+|\hla{1}-\hla{2}-(\hla{1}-\hla{3})v_{1}^{2}|
-2\ip{x}{v}^{2}\label{2mu}
\end{equation}
To describe the absolute minimum of the function (\ref{2mu}), we
need to consider auxiliary functions. The complexity of formulae
below mainly follows from the presence of the absolute value in
(\ref{2mu}). Let us define functions $p(\te)$ and $r(\te)$, as
follows. The common domain of $p(\te)$ and $r(\te)$  is the closed
interval $[-\pi,\pi]$. To define these functions we need the gap
angle $\vf_{\mr{gap}}\in (0,\pi/2)$, determined by the equation
$$
\cos \vf_{\mr{gap}}=\sqrt{\frac{\hla{1}-\hla{2}}{\hla{1}-\hla{3}}}
$$
Moreover, let $\sigma\,:\, [-\pi,\pi]\to [-1,1]$ be any function such
that
$$
\sigma(\te)=\W{1}{\te\in (-\pi,-\pi/2)\cup (0,\pi/2)}{-1}{\te\in
(-\pi/2,0)\cup (\pi/2,\pi)}
$$
and $\cO_{\mr{gap}}$ be the open set
$$
\cO_{\mr{gap}}=(-\pi+\vf_{\mr{gap}},-\vf_{\mr{gap}})\cup
(\vf_{\mr{gap}},\,\pi-\vf_{\mr{gap}})
$$
Now we are ready to define
$$
p(\te)=\W{1}{\te\in
\cO_{\mr{gap}}}{\cos^{2}(\te-\sigma(\te)\,\vf_{\mr{gap}})}{\te\in
[-\pi,\pi]\setminus \cO_{\mr{gap}}}
$$
and
$$
r(\te)=\W{\cos^{2}(\te-\sigma(\te)\,\vf_{\mr{gap}})}{\te\in
\cO_{\mr{gap}}}{1}{\te\in [-\pi,\pi]\setminus \cO_{\mr{gap}}}
$$
Finally, let $\te_{\ast}$ and $\te_{\ast\ast}$ be the angles defined
below in two steps. First, let $x_{1}x_{3}\neq 0$. Define
\begin{equation}
\cos
\te_{\ast}=\frac{\sqrt{2}|x_{1}x_{3}|}{\sqrt{N_{\ast}(N_{\ast}-\hla{1}+\hla{3}-x_{1}^{2}+x_{3}^{2})}},
\quad
\sin\te_{\ast}=-\frac{\sqrt{2}x_{1}x_{3}}{\sqrt{N_{\ast}(N_{\ast}+\hla{1}-\hla{3}+x_{1}^{2}-x_{3}^{2})}}
\label{testar}
\end{equation}
where
$$
N_{\ast}=\sqrt{(\hla{1}-\hla{3}+||x||^{2}-x_{2}^{2})^{2}-4(\hla{1}-\hla{3})x_{3}^{2}}
$$
and
\begin{equation}
\cos
\te_{\ast\ast}=\frac{\sqrt{2}|x_{1}x_{3}|}{\sqrt{N_{\ast\ast}(N_{\ast\ast}-x_{1}^{2}+x_{3}^{2})}},\quad
\sin\te_{\ast\ast}=-\frac{\sqrt{2}\,x_{1}x_{3}}{\sqrt{N_{\ast\ast}(N_{\ast\ast}+x_{1}^{2}-x_{3}^{2})}}\label{testar2}
\end{equation}
with
$$
N_{\ast\ast}=||x||^{2}-x_{2}^{2}
$$
If $x_{1}x_{3}=0$ we pass to the one - sided limits $x_{1}\to 0^{\pm}$ or $ x_{3}\to 0^{\pm}$ (or both) in the formulas (\ref{testar}) and (\ref{testar2}). The
left and right limits can only differ by sign, but this has no
importance as the functions $p(\te)$ and $r(\te)$ give the same
result in these situations.
\par To obtain the upper bound for quantum discord, define
\begin{equation}
\mu_{\ast}=\frac{1}{2}\left(\hla{1}+\hla{3}+||x||^{2}+x_{2}^{2}+N_{\ast}\,(1-2p(\te_{\ast})\right)\label{muast}
\end{equation}
and
\begin{equation}
\mu_{\ast\ast}=\hla{2}+||x||^{2}-(||x||^{2}-x_{2}^{2})r(\te_{\ast\ast})\label{muast2}
\end{equation}
Then one can show that
$$
D_{1}(\ro)\leq \sqrt{\mr{min}\,\{\mu_{\ast},\, \mu_{\ast\ast}\}}
$$
The above analysis leads to the following result.
\begin{thm}
In the case of non-degenerate spectrum of the matrix
$KK^{\mr{T}}-W_{x}$, the value of trace - norm geometric discord is
given by
$$
D_{1}(\ro)=\sqrt{\min \{d_{1},\,d_{2},\,d_{3}\}}
$$
where
$$
d_{1}=\hla{1}+x_{1}^{2},\quad d_{2}=\frac{1}{2}\left(\hla{1}+\hla{2}+||x||^{2}+x_{3}^{2}-\sqrt{(\hla{1}-\hla{2}+||x||^{2}-x_{3}^{2})^{2}-4(\hla{1}-\hla{2})x_{2}^{2}}\right),\quad
d_{3}=\min\{\mu_{\ast},\,\mu_{\ast \ast}\}
$$
with $\mu_{\ast},\; \mu_{\ast \ast}$ given by (\ref{muast}) and
(\ref{muast2}) respectively.
\end{thm}
\subsection{Degenerate spectrum of $L_{-}$}
When there is any degeneracy in the spectrum of $L_{-}$, the analysis simplifies significantly. From the above discussion we obtain:
\begin{thm}
1. Let $\hla{1}=\hla{2}\geq \hla{3}$, then
$$
D_{1}(\ro)=\sqrt{\hla{2}}
$$
2. Let $\hla{1}>\hla{2}=\hla{3}$, then
$$
D_{1}(\ro)=\frac{1}{\sqrt{2}}\,\sqrt{\hla{1}+\hla{2}+||x||^{2}-\sqrt{(\hla{1}-\hla{2}+||x||^{2})^{2}+4\,(\hla{1}-\hla{2})x_{1}^{2}}}
$$
\end{thm}
\section{Examples}
In this section we present the applications of Theorems VI.1 and VI.2 to the cases already known in the literature as well as to the new examples of states. But we start with considering again, from the wider perspective,  the results obtained in Sections IV and V.D.
\subsection{The class $\fE_{0}$}
In this case the correlation matrix is of the form $K=t\,V_{0}$, where $V_{0}$ is orthogonal and the Bloch vectors $x$ and $y$ are arbitrary. Thus
$$
L_{-}=t^{2}\, \I_{3}-||x||^{2}\,P_{x}
$$
and
$$
\mr{spect}\,L_{-}=\{t^{2},\, t^{2},\, t^{2}-||x||^{2}\}
$$
So we can apply Theorem VI.2 to obtain
$$
D_{1}(\ro)=|t|
$$
\subsection{The class of MMM states}
In this case Bloch vectors $x$ and $y$ are zero vectors, so
$$
L_{-}=KK^{\mr{T}}
$$
Let $\las{1}\geq\las{2}\geq\las{3}$ be the eigenvalues of $KK^{\mr{T}}$. When the eigenvalues are degenerate, we apply Theorem VI.2 and obtain
$D_{1}(\ro)=\sqrt{\las{2}}$.
In the non-degenerate case we must apply Theorem VI.1 and the computation is more involved. Observe that in the case of MMM states
$$
d_{1}=\las{1},\quad d_{2}=\las{2},\quad
d_{3}=\min \{\mu_{\ast},\, \mu_{\ast\ast}\}
$$
where obviously $\mu_{\ast\ast}=\las{2}$. To show that $\mu_{\ast}$ also equals to $\las{2}$ one checks that $\te_{\ast}=0$. Then
$$
p(\te_{\ast})=\cos^{2}\vf_{\mr{gap}}=\frac{\las{1}-\las{2}}{\las{1}-\las{3}}
$$
so by (\ref{muast}),  $\mu_{\ast}=\las{2}$. Thus
$$
D_{1}(\ro)=\mr{min}\, \{d_{1},\, d_{2},\, d_{3}\}=\sqrt{\las{2}}
$$
\subsection{Pure states}
 It is well known that up to the local equivalence, every pure state has the form
\begin{equation}
\ro_{N}= \frac{1}{2}\begin{pmatrix} 1+\sqrt{1-N^{2}}&0&0&N\\
0&0&0&0\\0&0&0&0\\N&0&0&1-\sqrt{1-N^{2}}\end{pmatrix},\quad N\in [0,1]\label{pureN}
\end{equation}
The corresponding correlation matrix and Bloch vector are given by
$$
K=\begin{pmatrix}N&\hspace*{2mm}0&0\\0&-N&0\\0&\hspace*{2mm}0&1\end{pmatrix},\quad x=\begin{pmatrix}0\\0\\\sqrt{1-N^{2}}\end{pmatrix}
$$
So
$$
L_{-}=\begin{pmatrix}N^{2}&0&0\\0&N^{2}&0\\0&0&N^{2}\end{pmatrix}
$$
an we can apply Theorem VI.2 to obtain
$$
D_{1}(\ro_{N})=N
$$
On the other hand, the measure of entanglement defined by negativity, gives the same value.
\subsection{X-states}
Two - qubit X - state has the X - shaped form
\begin{equation}
\ro_{X}=\begin{pmatrix}\ro_{11}&0&0&\ro_{14}\\
0&\ro_{22}&\ro_{23}&0\\
0&\ro_{32}&\ro_{33}&0\\
\ro_{41}&0&0&\ro_{44}\end{pmatrix}\label{Xstate}
\end{equation}
We can always take off - diagonal matrix elements $\ro_{14}$ and
$\ro_{23}$ to be positive \cite{cicca}. Then the correlation matrix
$K$ is diagonal and the Bloch vector $x$ has only third component
non-zero. So the matrix $KK^{\mr{T}}$ is  diagonal and commutes
with the operator $W_{x}$. It follows that the matrix $L_{-}$ has
also diagonal form and the basis of eigenvector of $L_{-}$ coincides
with the canonical basis. Since the spectrum of $L_{-}$ is in
general non-degenerate, to find the value of $D_{1}(\ro_{X})$ we
apply the Theorem VI.1. Let us start with computing $\mu_{\ast}$ and
$\mu_{\ast\ast}$. Notice that  this case when $x_{1}x_{3}=0$ as
$x_{1}=x_{2}=0$. Let
$$
n_{\ast}=\hla{1}-\hla{3}+x_{1}^{2}-x_{3}^{2}
$$
then
$$
\cos\te =\frac{1}{\sqrt{2}}\sqrt{1+\frac{n_{\ast}}{\sqrt{n_{\ast}^{2}+4x_{1}^{2}x_{3}^{2}}}}\quad\text{and}\quad \cos\te_{\ast}=\lim\limits_{x_{1}\to 0}\cos \te
$$
We can choose the sign of $x_{1}$ such that $\sin \te >0$, so the angles are in the interval $[0, \pi/2]$.
Computing this limit we obtain: \\[2mm]
$\bullet$ for $\hla{1}>\hla{3}+||x||^{2},\; \te_{\ast}=0$, then
$N_{\ast}=\hla{1}-\hla{3}-||x||^{2},\;\sigma (\te_{\ast})=\pm 1$\\
$\bullet$ for $\hla{1}=\hla{3}+||x||^{2},\; \te_{\ast}=\pi/4$, then $N_{\ast}=0$\\
$\bullet$ for $\hla{1}<\hla{3}+||x||^{2},\; \te_{\ast}=\pi/2$, then $N_{\ast}=\hla{3}+||x||^{2}-\hla{1},\; \sigma (\te_{\ast})=\pm 1$.\\[2mm]
Let $\te_{\ast}=0$, then $p(0)=\cos^{2}(0\pm \vf_{\mr{gap}})=\cos^{2}(\vf_{\mr{gap}})$, so
$$
\mu_{\ast}=\hla{2}+||x||^{2}\,\frac{\hla{1}-\hla{2}}{\hla{1}-\hla{3}}
$$
If $\te_{\ast}=\pi/4$, then $\hla{1}=\hla{3}+||x||^{2}$ and
$$
\mu_{\ast}=\hla{1}
$$
Finally, if $\te_{\ast}=\pi/2$, then $p(\te_{\ast})=1$ and again
$$
\mu_{\ast}=\hla{1}
$$
Summarizing
$$
\mu_{\ast}=\W{\hla{1}}{\hla{1}\leq \hla{3}+||x||^{2}}{\frac{\DS \hla{1}(\hla{2}+||x||^{2})-\hla{2}(\hla{3}+||x||^{2})}{\DS \hla{1}-\hla{3}}}{\hla{1}>\hla{3}+||x||^{2}}
$$
Now consider $\mu_{\ast\ast}$. Observe that $\cos\te_{\ast\ast}=\lim\limits_{x_{1}\to 0}\cos \te=0$. So $\te_{\ast\ast}=\pi/2$ and
$$
r(\te_{\ast\ast})= \cos^{2}(\pi/2 -\sigma(\te_{\ast\ast})\vf_{\mr{gap}})=\sin^{2}\vf_{\mr{gap}}
$$
Thus
$$
\mu_{\ast\ast}=\hla{2}+||x||^{2}-||x||^{2}\,\sin^{2}\vf_{\mr{gap}}=\hla{2}+||x||^{2}-||x||^{2}\,\frac{\hla{2}-\hla{3}}{\hla{1}-\hla{3}}=\hla{2}+||x||^{2}\, \frac{\hla{1}-\hla{2}}{\hla{1}-\hla{3}}
$$
Next we see that $d_{1}=\hla{1}$ and $d_{2}=\hla{2}+||x||^{2}$. Finally one can check that
$$
\min\,\{d_{1},d_{2},d_{3}\}=\hla{1}\quad\text{for}\quad \hla{1}\leq \hla{3}+||x||^{2}
$$
and
$$
\min\,\{d_{1},d_{2},d_{3}\}=\mu_{\ast}=\mu_{\ast\ast}\quad\text{for}\quad \hla{1}>\hla{3}+||x||^{2}
$$
so
$$
D_{1}(\ro_{X})=\W{\hla{1}}{\hla{1}\leq \hla{3}+||x||^{2}}{\frac{\DS \hla{1}(\hla{2}+||x||^{2})-\hla{2}(\hla{3}+||x||^{2})}{\DS \hla{1}-\hla{3}}}{\hla{1}>\hla{3}+||x||^{2}}
$$
and this formula is in the full accordance with  the results presented in \cite{cicca}.
\subsection{The class containing quantum-classical states}
Now we apply our results to the class of states with arbitrary Bloch
vector $x$ and correlation matrix having only one non-zero
singular value. Such class in particular contains quantum -
classical states \cite{cicca}. We can always assume that
$$
K=\begin{pmatrix}\kappa&0&0\\0&0&0\\0&0&0\end{pmatrix},\quad
x=\begin{pmatrix}x_{1}\\0\\x_{3}\end{pmatrix}
$$
One can check that possible critical points lay on the big circle
$v_{2}=0$ or are given by $(0,\pm 1,0)$. By a direct inspection one
gets that the minimal value can be achieved only on the circle
$v_{2}=0$. So we are left with one dimensional minimization problem.
This can be done by a direct analysis (see \cite{cicca}), but we
apply our general result. To find value of $D_{1}(\rho)$ we need
only $\mu_{\ast}$ and $\mu_{\ast\ast}$ which take the form
$$
\mu_{\ast}=\kappa^{2}\, (1-p(\theta_{\ast})),\quad
\mu_{\ast\ast}=||x||^{2}\,(1-r(\theta_{\ast\ast}))
$$
with $\theta_{\ast}\in (\vf_{\mr{gap}},\vf_{\mr{gap}})$ and
$\theta_{\ast\ast}\in (-\pi/2,-\vf_{\mr{gap}})\cup
(\vf_{\mr{gap}},\pi/2)$. After a direct but quite lengthy
computations one  finds that the absolute minimum is given by
$\mu_{\ast}$ and the value of discord reads
$$
D_{1}(\ro)=\frac{\DS |\kappa|\,|x_{3}|}{\DS
\sqrt{(\kappa+x_{1})^{2}+x_{3}^{2}}}
$$
 Notice that this formula  is in a full agreement with the result
obtained in \cite{cicca}.
\subsection{Beyond the X-states}
Consider the following two-parameter family of states
\begin{equation}
\ro_{\g,a}=\frac{1}{4}\begin{pmatrix}1+a(1+w_{2})&-iaz&-iaz&a(2-w_{1})\\iaz&1+a(1-w_{2})&a(2+w1)&iaz\\iaz&a(2+w_{1})&1-a(1+w_{2})&iaz\\
a(2-w_{1})&-iaz&-iaz&1-a(1-w_{2})\end{pmatrix}\label{roag}
\end{equation}
where for $\fala{\g}=\sqrt{1+16\g^{2}}$ we define
\begin{equation*}
\begin{split}
&w_{1}=\frac{\sqrt{7-\fala{\g}}\hspace*{1mm}(\fala{\g}-1)+\sqrt{7+\fala{\g}}\hspace*{1mm}(\fala{\g}+1)}{2\sqrt{2}\:\fala{\g}}\\[2mm]
&w_{2}=\frac{\sqrt{7-\fala{\g}}\hspace*{1mm}(\fala{\g}+1)+\sqrt{7+\fala{\g}}\hspace*{1mm}(\fala{\g}-1)}{2\sqrt{2}\:\fala{\g}}\\[2mm]
&z=\frac{\sqrt{7+\fala{\g}}-\sqrt{7-\fala{\g}}}{\fala{\g}}\;\sqrt{2}\,\g
\end{split}
\end{equation*}
One can check that (\ref{roag}) defines a state if $|\g|\leq \sqrt{3}$ and $a$ is the function of $\g$, known only numerically. The corresponding
correlation matrix and Bloch vector are as follows
$$
K=a\, \begin{pmatrix}2&0&0\\0&w_{1}&z\\0&z&w_{2}\end{pmatrix},\quad x=\begin{pmatrix}0\\0\\a\end{pmatrix}
$$
so the parameter $a$ is equal to the norm $||x||$ of the Bloch vector. Interesting property of the above correlation matrix is that
$$
KK^{\mr{T}}= ||x||^{2}\,\begin{pmatrix}4&0&0\\0&4&2\g\\0&2\g&3\end{pmatrix}
$$
so the matrix $L_{-}$ is simple
\begin{equation}
L_{-}=||x||^{2}\,\begin{pmatrix}4&0&0\\0&4&2\g\\0&2\g&2\end{pmatrix}\label{Lmg}
\end{equation}
and the corresponding ordered eigenvalues are given in the explicit way
\begin{equation}
\hla{1}=||x||^{2}\,\left(3+\sqrt{1+4\g^{2}}\right),\quad \hla{2}=4\,||x||^{2},\quad \hla{3}=||x||^{2}\,\left(3-\sqrt{1+4\g^{2}}\right)
\end{equation}
Since the spectrum of $L_{-}$ is non-degenerate, to find the value of $D_{1}(\ro_{\g,a})$, we must apply the Theorem VI.1. First notice that the Bloch vector in the basis of eigenvectors of $L_{-}$ reads
$$
x=||x||\,\begin{pmatrix}\frac{\DS 2\g}{\DS \sqrt{2\left(1+4\g^{2}+\sqrt{1+4\g^{2}}\right)}}\\[2mm] 0\\[2mm]
\frac{\DS -2\g}{\DS \sqrt{2\left(1+4\g^{2}-\sqrt{1+4\g^{2}}\right)}}\end{pmatrix}
$$
Then
$$
\cos \vf_{\mr{gap}}=\frac{\sqrt{2}|\g|}{\sqrt{1+4\g^{2}+\sqrt{1+4\g^{2}}}},\quad \sin\vf_{\mr{gap}}=\frac{\sqrt{2}|\g|}{\sqrt{1+4\g^{2}-\sqrt{1+4\g^{2}}}}
$$
Since $N_{\ast\ast}=||x||^{2}$
$$
\cos\te_{\ast\ast}=\frac{|x_{1}|}{||x||}=\frac{\sqrt{2}|\g|}{\sqrt{1+4\g^{2}+\sqrt{1+4\g^{2}}}}
$$
and
$$
\sin\te_{\ast\ast}=-\frac{x_{3}\mr{sign}(x_{1})}{||x||}=\frac{\sqrt{2}|\g|}{\sqrt{1+4\g^{2}-\sqrt{1+4\g^{2}}}}
$$
So $\te_{\ast\ast}\in (0,\pi/2)$ and $\te_{\ast\ast}=\vf_{\mr{gap}}$. It follows that $r(\te_{\ast\ast})=1$ and
$$
\mu_{\ast\ast}=\hla{2}+||x||^{2}-||x||^{2}=\hla{2}=4\,||x||^{2}
$$
Next, since $N_{\ast}=||x||^{2}\sqrt{1+16\,\g^{2}}$ we obtain that
$$
p(\te_{\ast})=\cos^{2} (\te_{\ast}-\vf_{\mr{gap}})=\frac{1}{2}\left(1-\frac{1}{\sqrt{1+16\,\g^{2}}}\right)=\frac{1}{2}\left(1-\frac{||x||^{2}}{N_{\ast}}\right)
$$
So applying (\ref{muast}) we obtain
$$
\mu_{\ast}=\frac{1}{2}\left(7\,||x||^{2}+N_{\ast}\left(1-(1-\frac{||x||^{2}}{N_{\ast}})\right)\right)=4\,||x||^{2}=\mu_{\ast\ast}
$$
On the other hand
$$
d_{1}=\hla{1}+x_{1}^{2}=||x||^{2}\left(3+\sqrt{1+16\g^{2}}\right)+x_{1}^{2}>4\,||x||^{2}
$$
and
$$
d_{2}=\frac{1}{2}\left(\hla{1}+\hla{2}+||x||^{2}+x_{3}^{2}-(\hla{1}-\hla{2})\right)=\hla{2}+x_{3}^{2}>\hla{2}
$$
So
$$
\min\,\{d_{1},d_{2},d_{3}\}=\mu_{\ast}=\mu_{\ast\ast}=4\,||x||^{2}
$$
and
$$
D_{1}(\ro_{\g a})=2\,||x||^{2}
$$
\section{Summary}
Let us now summarize the results reported in the article. First, we have obtained new compact and elegant formula for
the trace norm of the disturbance $S(\cM)$ of the general two-qubit state. We want to emphasize, that to obtain  this
result we have used  the qubit structure of subsystems in the crucial way. In the general qudit case we were able only to
find simplified formula for the square of the modulus of disturbance and only for the  special classes of states \cite{LFJ}.
It seems that the computation of  trace norm of $S(\cM)$ in general case is a highly non-trivial or even not possible at all.
Then, we have applied the formula for $||S(\cM)||_{1}$ to compute the value of geometric discord in the situation when the
minimization procedure is not needed at all or is straightforward. This is the case of  the class of states with arbitrary
Bloch vectors and the correlation matrix proportional to some orthogonal matrix in three dimensions i.e. $K=t\, V_{0}$.
In particular, for such states with zero Bloch vectors, $||S(\cM)||_{1}=|t|$ and $D_{1}(\ro)=|t|$. The presence of
non-zero Bloch vectors only slightly changes the trace norm and the minimization can be done in  elementary way.
This again must be contrasted with the general qudit case, where the similar results are valid only for vanishing Bloch
vectors and the classes of Werner and isotropic states \cite{LFJ}. As a first step towards the computation of
$D_{1}(\ro)$ in general case, we considered critical points of the mapping $\cM\to ||S(\cM)||_{1}$. To solve this
problem, we  introduce the function $\mathbf{g}$ defined on the unit sphere $\Sp^{2}\subset \R^{3}$ with values in $\R^{2}$. Any critical point of the mapping $\cM\to ||S(\cM)||_{1}$ is a critical point of the function $\mathbf{g}$ and the properties of this function can be used to isolate singular and smooth critical points. It is remarkable that the information about critical points is encoded in the spectrum of the matrix $L_{-}=KK^{\mr{T}}-W_{x}$. One of the most important results of our study says that when this spectrum is non-degenerate, the critical points are located on big circles in $\Sp^{2}$. It means that the problem of finding geometric discord reduces to minimization of the quadratic form restricted to big circles. The complete solution of this problem is rather involved and it simplifies significantly, when any degeneracy of the spectrum of $L_{-}$ is present.

We apply the Theorems VI.1 and VI.2  to the cases already discussed in the literature (Bell diagonal, X-shaped  and 
quantum-classical states) as well as new examples beyond the class of X-states. It is also worth to emphasize that in particular cases, methods based on  partial results can be more efficient then the direct application of the general result, what is explicitly shown by the discussions of MMM states in Sections V.D and VII.B.

Having the complete answer for the two-qubit system we are able to
estimate possible obstacles in finding general explicit formula for
the trace-distance quantum discord for qudits. As we have already
stressed, the geometry  of the two-qubit quantum state space is
exceptional and the set of geometrical tools which is at our
disposal is very reach. Let us recall here  the most effective one,
that any correlation matrix can be diagonalized by means of local
transformations, due to the accordance of the dimensions of relevant
spaces and groups i.e. $\mathrm{SO(3)}\simeq \mathrm{Adj(SU(2))}$.
For the next higher level system of two-qutrits $d=3$ we have that
$\mathrm{Adj(SU(3)))}$ is a small subgroup of $\mathrm{SO(8)}$, too
small to allow diagonalization of a generic two-qutrit correlation matrix by
means of local transformations. For higher dimensional cases this
effects only aggravates, what means that using local orbit of the
set of diagonal correlation matrices we can cover only a part of the
quantum state space. The solution of the minimization problem for
some  families of two-qudit states is given in our previous works \cite{LFJ, jfl-pla}.

The strict  analytical general formulas characterising  quantum correlations valid for all or sufficiently wide
class of quantum states of compound systems, in general are not known and only selected cases are described in
the literature. Such explicit knowledge would be of value that hardly can be overestimated, as it gives insight into the
structure of  the quantum states, but on the other hand due to the complexity of this space, computation of
physically important measures of quantum correlations is so hard task. We hope that results obtained in the present paper
will fill the gap in the knowledge on measures of quantum correlations of compound systems for the most fundamental  physical
system of two qubits.


\begin{thebibliography}{99}
\bibitem{Modi} Modi, K., A. Brodutch, H. Cable, T. Paterek, V. Vedral, Rev. Mod. Phys. \textbf{ 84}, 1655 (2012).
\bibitem{Adesso} Streltsov, A., G. Adesso, M. B. Plenio, Rev. Mod. Phys. \textbf{89}, 041003 (2017).
\bibitem{HHWPZF}  Ming-Liang Hu, Xueyuan Hu, Jieci Wang, Yi Peng, Yu-Ran Zhang, Heng Fan, Phys. Rep. 726-764 (2018).
\bibitem{LFJ} P. {\L}ugiewicz, A. Frydryszak and L. Jak\'obczyk, J. Phys. A: Math. Theor. \textbf{50}, 245301 (2017).
\bibitem{GQD} B. Daki\'{c}, V. Vedral, {C}. Brukner, Phys.Rev. Lett.  \textbf{105}, 190502 (2010).
\bibitem{lj-hs} L. Jak\'obczyk, Phys. Lett. A \textbf{378}, 3248 (2014).
\bibitem{Na} T. Nakano, M. Piani and G. Adesso, Phys. Rev. A \textbf{88}, 012117 (2013).
\bibitem{RSI} W. Roga, D. Spehner and F. Illuminati, J. Phys. A: Math. Theor. \textbf{49}, 235301 (2016).
\bibitem{Paula}  F.M. Paula, T.R. de Oliveira and M.S. Sarandy, Phys. Rev. A \textbf{87}, 064101 (2013).
\bibitem{cicca} F. Ciccarello, T. Tufarelli, V. Giovannetti, New J. Phys \textbf{16}, 013038 (2014).
\bibitem{MM} H. Mehri - Dehnavi, B. Mizra, H. Mohammadzadeh and R. Rahimi, Ann. Phys. \textbf{326}, 1320 (2011).
\bibitem{jfl-pla} L. Jak\'{o}bczyk, A. Frydryszak, P {\L}ugiewicz, Phys. Lett. A \textbf{380}, 1535-1541 (2016).
\end{thebibliography}
\end{document}